\documentclass[12pt]{article}

\usepackage{amsmath}
\usepackage{amssymb}
\usepackage{rotating}
\usepackage{graphicx}

\usepackage{color} 

\usepackage[margin=1in]{geometry}
\usepackage{abstract}

\usepackage[labelfont=bf,labelsep=period,justification=raggedright]{caption}

\makeatletter
\renewcommand{\@biblabel}[1]{\quad#1.}
\makeatother

\newcommand{\comment}[1]{}

\date{}

\begin{document}

\title{\vspace*{-1.5cm}Transfer Functions for Protein Signal Transduction:
Application to a Model of Striatal Neural Plasticity}

\author{Gabriele Scheler\\
Carl Correns Foundation for Mathematical Biology\\
Mountain View, Ca, USA}

\maketitle

\begin{abstract}
\raggedright
We present a novel formulation for biochemical reaction networks in the context of protein signal transduction. The model consists of input-output transfer functions, which are derived from differential equations, using stable equilibria. We select a set of 'source' species, which are interpreted as input signals. Signals are transmitted to all other species in the system (the 'target' species) with a specific delay and with a specific transmission strength. The delay is computed as the maximal reaction time until a stable equilibrium for the target species is reached, in the context of all other reactions in the system. The transmission strength is the concentration change of the target species.  The computed input-output transfer functions can be stored in a matrix, fitted with parameters, and even recalled to build dynamical models on the basis of state changes. By separating the temporal and the magnitudinal domain we can greatly simplify the computational model, circumventing typical problems of complex dynamical systems.
The transfer function transformation of biochemical reaction systems can be applied to mass-action kinetic models of signal transduction. 
The paper shows that this approach yields significant novel insights while remaining a fully testable and executable dynamical model for signal transduction.    
In particular we can deconstruct the complex system into local transfer functions between individual species. 
As an example, 
we examine modularity 
and signal integration using a published model of striatal neural plasticity.
The modularizations that emerge correspond to a known biological distinction between calcium-dependent and cAMP-dependent pathways. 
Remarkably, we found that overall interconnectedness depends on the magnitude of inputs, with higher connectivity at low input concentrations and significant modularization at moderate to high input concentrations. This general result, which directly follows from the properties of individual transfer functions, contradicts notions of ubiquitous complexity by showing input-dependent signal transmission inactivation.
\end{abstract}

\section*{Introduction}

Biochemical reaction systems are usually conceptualized as dynamical systems - systems that evolve in continuous time and may or may not receive additional input to the system.
Mathematically, this can be expressed by sets of ordinary differential equations (ODE),
such that rates of concentration changes correspond to mass-action kinetic parameters 
\cite{BhallaandIyengar1999,Gilbert2006}. 
In this paper we use existing mass-action dynamical systems to propose an alternate or additional framework for modeling and interpretation of biochemical reaction systems.
We provide an algebraic analysis of biochemical reaction systems as a matrix of concentrations for all species,
given certain input concentrations. These concentrations correspond to steady-state amounts which are reached after a 
delay time, and the delay times can be measured by the system as well.

We use an arbitrary published model \cite{Lindskog2006} as an example for a ODE dynamical model of biochemical reactions. 
The model simulates intracellular signal transduction from receptor binding to molecular targets in different cellular compartments, as an important component in the long-term regulation of protein expression implied in neural and synaptic plasticity. In striatal neurons, both a calcium-dependent pathway and a cAMP-dependent pathway are activated during the initiation of neural plasticity by NMDA/AMPA receptors and neuromodulator receptors such as dopamine D1 receptors 
\cite{Svenningsson2004,Cepeda2006,Yger2011}. Their effects and the integration of signaling on common targets such as kinases and phosphatases have been the subject of a number of computational models 
\cite{Kotaleski2010,Stefan2008a,Kim2010,Pepke2010,Manninen2010}. In particular, the role of the DARPP-32 protein in striatal neurons in determining the outcome of membrane signaling has been modeled by different groups, based on a common set of experimental data
\cite{Nakano2010,Fernandez2005,Lindskog2006}, 
cf.\ \cite{LeNovere2008,Qi2008,Qi2010,OliveiraRF2012,Barbano2007}. Many similar models
 \cite{Li2010} have been developed in the last 10-12 years in different areas of biology. Models  with dozens or more of species have up to a 100 or  more equations and are consequently complex and difficult to understand as continuous dynamical systems 
\cite{Hughey2010}. A transformation into a matrix-based formulation of input-output functions, even at the cost of a loss of fast dynamical modeling, promises considerable gain of insight and access to a different set of mathematical tools.
Simple mass-action kinetic models may be criticized for disregarding the real complexity of spatio-temporal molecular interactions.
Some alternatives use spatial grids and stochastic versions of biochemical reactions to capture this complexity \cite{Wu2011,Erhard2008}.
However, certain 
variations, such as compartmental modeling with diffusion, altered kinetics for anchored proteins, or employing molecular kinetics as the basis for binding constants may be employed within the mass-action kinetic framework to achieve better correspondence with the biological reality. These variations can be directly transferred to the proposed model as well.

In our approach, we identify input nodes, and then pre-compute the outcomes for all internal species (target species) in response to biological meaningful ranges and combinations of inputs. This allows to analyze a biochemical reaction system under all possible input conditions. The analysis can be done for arbitrary ODE models \cite{Li2010}, provided minimal requirements on conservation properties are realized (cf.~section ``Methods'', \cite{Deng2011}).  
The results are stored as vectors or matrices ('systemic protein signaling functions' (psfs)) and can be fitted with functional parameters.
It is an important aspect of the model that computations are done systemically. 
In section ``Elementary Biochemical Reactions'',
source-target interactions are first analyzed in isolation ('elementary psfs'). They all constitute hyperbolic saturation functions, therefore rate parameters can be uniformly translated into functional parameters for signal transmission strength.  But in a systemic context, source-target interactions change because of additional influences on the species from other equations in the system 
(section ``Systemic PSF Analysis'').
Therefore a fitted systemic psf from A to B is different from the elementary psf. The pre-computed, systemic psfs may be used to create state-change simulation models, i.e. discrete-time models, which can be compared with continuous ODE models 
(section ``Systemic Delay and State-change Dynamics'').
What is significant and novel about our analysis is that we can extract systemic
transfer functions from the complex system, and thereby dissect the system into parts.  We can analyze the transmission properties of individual species, compare their minimum and maximum values, and the functional shape of their transmission strength. Specifically, we can show under which circumstances a link is functional, i.e. actually transmits information 
(section ``Computing Input-dependent Modularity'').

The analysis has a number of restrictions. An important restriction is that our model does not allow for analysis of fast interactions below the resolution of settling into steady-state. 
The requirement of conservation of mass guarantees that for each input concentrations will eventually settle to some equilibrium value, but due to the prevalence of feedback interactions, they may still produce transients or dampened oscillations. This means that fast fluctuations of input will not be adequately simulated using pre-computed psf functions alone. It is then necessary to refer to the underlying ODE model. The model is most suitable for studying disease states, pharmacological interventions, genetic manipulations, miRNA interference, or any system conditions which fundamentally alter the presence or concentration of molecular species. These conditions may then be tested either in steady-state or with a sequence of sufficiently slow input constellations.
A second restriction is that the model inherits parameter uncertainty from mass-action kinetic models. 
These parameters are derived from experimental measurements, but typically with a high degree of uncertainty \cite{Bandara2009a}. Our analysis 
offers a clear distinction between elementary and systemic functional parameters and explains why experimental measurements are so highly dependent on the systemic context. In this paper, we have treated elementary parameters as given by the underlying model \cite{Lindskog2006}. In principle, systemic functional interactions can be measured experimentally, and in the model each interaction can be adjusted separately. This may offer a novel theoretical approach towards finding adequate elementary rate parameters.

\section*{Materials and Methods}

\subsection*{System Definition}
\label{methods-1}

A biochemical reaction system formulation for signal transduction contains two different types of reactions:  
\begin{enumerate}
\item
complex formation\\
$[A]+[B] \leftrightarrow [AB]$
\item
enzymatic reactions\\ 
$[A]+[E]\leftrightarrow [AE] \rightarrow [A*]+E$.
\end{enumerate}
The system has concentrations for species A, B and E (A*, AE and AB can be calculated), a set of 
kinetic rate parameters $k_{on}$ and $k_{\mbox{\scriptsize\em off}}$ for the forward and backward binding reactions in complex formation,
and $k_{cat}$ for the rate of enzymatic production. The equational structure, the kinetic parameters and the initial concentrations  of the model \cite{Lindskog2006} are reproduced in Tables~\ref{kimbl1}, \ref{kimbl2}, \ref{kimbl3}, \ref{kimbl4}, with slight modifications: Equation 40 was added to 'close the loop' from AMP to ATP and thus provide for conservation of all molecules in the system. Conservation of mass is necessary in order for all species to reach equilibrium.
It means that for any forward reaction there needs to be a reverse reaction, such that any species receives both input and output ('weakly reversible system', cf.  \cite{Deng2011}). This implies that pure loss reactions, like endocytosis or diffusion across the cell membrane (secretion) cannot adequately be modeled with this system, unless balancing reactions are added which make up for the loss. E.g., for endocytosis of a ligand-bound receptor, both the ligand and the receptor, possibly independently, have to be recycled, which means that bridging equations for receptor-ligand dissolution in the endocytosed state, and input rates for receptors and ligands have to be added. 
Secondly, the species PP1 and its interactions (4 equations) were left out, since they contain a complex which dissolves into its 3 components in one step: this would require a, fairly trivial, addition to the current psf system implementation. 
The system can be depicted as a bipartite graph with nodes for species and nodes for reactions. 

\begin{table}[!ht]
{\footnotesize
\begin{tabular}{rlllll}
&& $k_{on}$&$k_{\mbox{\scriptsize\em off}}$&$k_{cat}$\\
1 & $ Da + D1R \leftrightarrow DaD1R$ & 0.00111 & 10 & \\
2 & $ DaD1R + Gabc \leftrightarrow DaD1RGabc$ & 0.0006 & 0.001 & \\
3 & $ Gabc + D1R \leftrightarrow GabcD1R$ & 6e-005 & 0.0003 & \\
4 & $ GabcD1R + Da \leftrightarrow DaD1RGabc$ & 0.00333 & 10 & \\
5 & $ DaD1RGabc \rightarrow DaD1R + GoaGTP + Gbc$ & & & 20\\
6 & $ GoaGTP \rightarrow GoaGDP$ & & & 10\\
7 & $ GoaGDP + Gbc \rightarrow Gabc$ & 100 &  & \\
8 & $ GoaGTP + AC5 \leftrightarrow AC5GoaGTP$ & 0.0385 & 50 & \\
9 & $ ATP + AC5GoaGTP \leftrightarrow AC5GoaGTP\_ATP \rightarrow$ &&&\\
& \hspace*{5cm}$cAMP + AC5GoaGTP$ & 0.000128 & 0.261 & 28.46\\
10 & $ AC5 + Ca \leftrightarrow AC5Ca$ & 0.001 & 0.9 & \\
11 & $ AC5Ca + GoaGTP \leftrightarrow AC5CaGoaGTP$ & 0.0192 & 25 & \\
12 & $ ATP + AC5CaGoaGTP \leftrightarrow AC5CaGoaGTP\_ATP \rightarrow$ &&& \\
&\hspace*{5cm}$cAMP + AC5CaGoaGTP$ & 6e-005 & 0.131 & 14.23\\
13 & $ PDE1 + Ca4CaM \leftrightarrow PDE1CaM$ & 0.1 & 1 & \\
14 & $ cAMP + PDE1CaM \leftrightarrow PDE1CaM\_cAMP \rightarrow AMP + PDE1CaM$ & 0.0046 & 44 & 11\\
15 & $ cAMP + PDE4 \leftrightarrow PDE4\_cAMP \rightarrow AMP + PDE4$ & 0.02 & 72 & 18\\
16 & $ PKA + cAMP \leftrightarrow PKAcAMP2$ & 2.6e-005 & 0.006 & \\
17 & $ PKAcAMP2 + cAMP \leftrightarrow PKAcAMP4$ & 3.46e-005 & 0.06 & \\
18 & $ PKAr + PKAc \leftrightarrow PKAcAMP4$ & 0.00102 & 0.0048 & \\
\end{tabular}
}
\caption{{Reactions in the cAMP pathway}}
\label{kimbl1}
\end{table}

\begin{table}[!ht]
\footnotesize
\begin{tabular}{rlllll}
&& $k_{on}$&$k_{\mbox{\scriptsize\em off}}$&$k_{cat}$\\
19 & $ Ca4CaM + PP2B \leftrightarrow PP2BCa4CaM$ & 1 & 0.3 & \\
20 & $ PP2BCa2CaM + Ca \leftrightarrow PP2BCa4CaM$ & 0.1 & 10 & \\
21 & $ CaM + PP2B \leftrightarrow PP2BCaM$ & 1 & 3 & \\
22 & $ Ca2CaM + PP2B \leftrightarrow PP2BCa2CaM$ & 1 & 0.3 & \\
23 & $ PP2BCaM + Ca \leftrightarrow PP2BCa2CaM$ & 0.006 & 0.91 & \\
24 & $ CaM + Ca \leftrightarrow Ca2CaM$ & 0.006 & 9.1 & \\
25 & $ Ca2CaM + Ca \leftrightarrow Ca4CaM$ & 0.1 & 1000 & \\
26 & $ Ca4CaM + CaMKII \leftrightarrow CaMKIICa4CaM$ & 0.00075 & 0.1 & \\
27 & $ CaMKIICa4CaM \rightarrow CaMKIIpCa4CaM$ & & & 0.005\\
28 & $ CaMKIIpCa4CaM \rightarrow CaMKIICa4CaM$ & & & 0.015\\
\end{tabular}
\caption{{Reactions in the Ca pathway}}
\label{kimbl2}
\end{table}

\begin{table}[!ht]
\footnotesize
\begin{tabular}{rlllll}
&& $k_{on}$&$k_{\mbox{\scriptsize\em off}}$&$k_{cat}$\\
29 & $ DARPP32 + PKAc \leftrightarrow DARPP32PKAc \rightarrow pThr34 + PKAc$ & 0.0027 & 8 & 2\\
30 & $ PP2A + PKAc \leftrightarrow PKAcPP2A \rightarrow PP2Ap + PKAc$ & 0.0025 & 0.3 & 0.1\\
31 & $ PP2Ap \rightarrow PP2A$ & & & 0.004\\
32 & $ pThr34 + PP2BCa4CaM \leftrightarrow pThr34PP2B \rightarrow$ &&&\\
&\hspace*{5cm}$DARPP32 + PP2BCa4CaM$ & 0.001 & 2 & 0.5\\
33 & $ pThr34 + PP2A \leftrightarrow pThr34PP2A \rightarrow DARPP32 + PP2A$ & 0.0001 & 2 & 0.5\\
34 & $ DARPP32 + Cdk5 \leftrightarrow DARPP32Cdk5 \rightarrow pThr75 + Cdk5$ & 0.00045 & 2 & 0.5\\
35 & $ pThr75 + PKAc \leftrightarrow pThr75PKc$ & 0.00037 & 1 & \\
36 & $ pThr75 + PP2Ap \leftrightarrow pThr75PP2Ap \rightarrow DARPP32 + PP2Ap$ & 0.0004 & 12 & 3\\
37 & $ pThr75 + PP2A \leftrightarrow pThr75PP2A \rightarrow DARPP32 + PP2A$ & 0.0001 & 6.4 & 1.6\\
38 & $ 1 PP2A + 4 Ca \leftrightarrow 1 PP2Ac$ & 7.72e-012 & 0.01 & \\
39 & $ pThr75 + PP2Ac \leftrightarrow pThr75PP2Ac \rightarrow DARPP32 + PP2Ac$ & 0.0004 & 12 & 3\\
40 & $ AMP \rightarrow ATP$ & & & 10\\
\end{tabular}
\caption{{DARPP-32 reactions }}
\label{kimbl3}
\end{table}

\begin{table}[!ht]
\begin{tabular}{llllllll}
$D1R$ & 500&$CaMKII$ & 20000&$PDE1$ & 4000&$PP1$ & 5000\\
$Gabc$ & 3000&$DARPP32$ & 50000&
$PDE4$ & 2000&$Cdk5$ & 1800\\
$AC5$ & 2500&$PP2A$ & 2000&$PKA$ & 1200&$Ca$ & 1000\\
$ATP$ & 2e+006&$PP2B$ & 4000&$CaM$ & 10000&$Da$ & 5000
\end{tabular}
\caption{{Initial Concentrations}}
\label{kimbl4}
\end{table}

\subsection*{PSF Analysis}

In order to set up source-target functions, we need to select input nodes from the available species nodes. In this example, we used $Da$  (dopamine as ligand for the D1 receptor) and $Ca$ (extracellular calcium that diffuses through ion channels in the membrane). 
We use input concentrations over a specified range (e.g., between 60nM and $5 \mu M$ for Da), sample over the range with e.~g. n=20 steps, and use the differential equation implementation of the system to calculate the output values for all species for each sampling step.
Because of the conservation of molecules, all species reach steady-state after a sufficient period of time.
We define steady-state pragmatically by relative change of less than 2\% over 100s. We also use the established terminology of EC10, EC50, EC90 etc. to indicate 10, 50 or 90 \% of steady-state concentration value. Additionally, we calculate the delay in reaching steady-state. 
We store input-target concentration mappings in a vector (single-input system), or a matrix (multiple-input system). 
We fit the vectors with hyperbolic or linear functions, using standard techniques in Matlab 
({\it fminsearch}, \cite{Lagarias98}). In this way we derive parameters which can be analyzed and used instead of the explicit vectors. (In this paper, the fitting is only done for single-input systems, multiple-input systems require different techniques.)
 
All information on source-target transfer functions for the complete, complex signaling system ('systemic psf') can be stored in a static data structure. For each species, it contains its concentration range, and for each reaction, it contains the parameters 
of the functional fit.  We gain the possibility to regard any species as source and any other species as target (they may be coupled by an arbitrary number of reactions) and obtain a systemic psf as the transfer function between them.
This representation allows to analyze the complex signaling system by its parts, i.e. as a set of matrices or vectors, which is the main achievement relative to the ODE dynamical system. In addition, dynamical simulation with appropriate update times may be realized by the psf representation alone, i.e. the psf simulation is not in itself atemporal, but only discrete and fairly slow.

We visualize the data structure as a bipartite graph, and label it with the calculated numeric values. Each species node is labeled with its attainable concentration range given the input range. For complex formation reactions, we show both $[A]\rightarrow[AB]$ and $[B]\rightarrow [AB]$.  For enzymatic reactions we show [E] as the source and [A*] as the target ($[E] \rightarrow [A*]$). 
The result is a labeled bipartite graph, called a 'weighted dynamic graph'.


\section*{Results}
\subsection*{Elementary Biochemical Reactions}
\label{elementary}
We want to represent 
a biochemical reaction by a time-independent signal transfer function, such that 
$y=f(x)$ for two species $x, y$. 
We do this by designating a source species x and then calculating the steady-state value for another species, the target species y, for any value of x, given the differential equations for the biochemical reaction. 
For complex formation \[[A]+[B] \leftrightarrow [AB]\] where the total concentrations for [A] and [B]
and kinetic rate parameters $k_{on}$, $k_{\mbox{\scriptsize\em off}}$ (with {\small $Kd ={koff\over{kon}}$}) are given, the differential equations are:
\[
dxdt(A) = k_{\mbox{\scriptsize\em off}} [AB] - k_{on}[A][B]\]\[
dxdt(B) = k_{\mbox{\scriptsize\em off}} [AB] - k_{on}[A][B]\]\[
dxdt(AB) = - k_{\mbox{\scriptsize\em off}} [AB] + k_{on}[A][B]\]
 
\begin{figure}[!ht] 
\centerline{\includegraphics[width=0.99\textwidth]{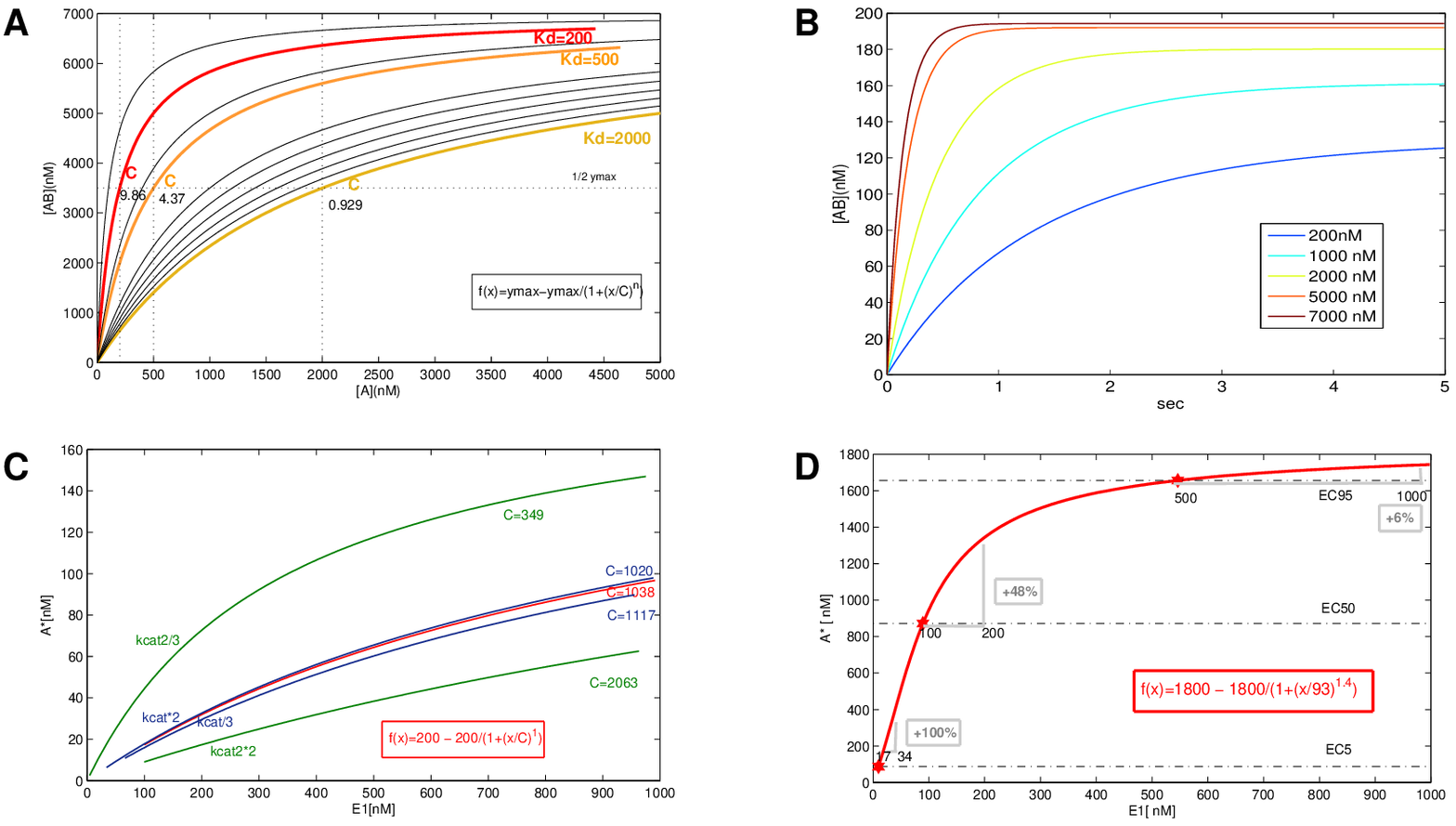}}
  \caption{
{\bf Properties of psf functions.}
{\bf A.} Complex formation:
Psfs were generated for $[A]+[B]\leftrightarrow [AB]$ with $Kd=200, 500, 2000$, $[A]_0=10nM-10 \mu M$; 
$[B]_0=7 \mu M$, and fitted with 
the saturating hyperbolic function shown in the figure. For an elementary reaction, $C=Kd$. All other curves were 
generated only by varying C, where $y_{max}=[B]_0$ and n=1. The slope at C indicates the signal transmission strength.
{\bf B.} Dynamics of complex formation: $[A]+[B]\leftrightarrow [AB]$ with $Kd=200$, $[B]_0=200nM$; $[A]_0=200nM-7 \mu M$. Increasing concentrations for the source species $[A]_0$ speeds up the reaction.
{\bf C.} Enzymatic Reactions with Reverse Reaction: 
Example reaction with $[A]_0$= 200, $k_{on}$=0.00026; $k_{\mbox{\scriptsize\em off}}$=1.5;
$k_{cat}$=30.4; 
$k_{cat2}$=0.26 (shown in red) and variations (shown in green, blue). 
Both the forward reaction ($k_{cat}$) and reverse reaction ($k_{cat2}$) parameters are varied by 30\%--200\%. Variability in signal transmission is expressed by the fitted parameter C (half-maximal activation) in a uniform way.
{\bf D.} Signal Transmission Strength: A 100\% increase of input yields diminishing increases at higher concentrations. Signals are shown at EC5, EC50 and EC95.
}
\label{fig:f1234}
\end{figure}

We may now calculate the concentration values $f(x)= y$ for a target species [AB] given a range of input values for $x$, e.g. the source species [A]. (Fig. \ref{fig:f1234}A).


In this way we separate the calculation of the signal response magnitude, i.~e.~the steady-state concentration, from the calculation of the time until a steady-state value is reached, the delay. 
For different $x$, $f(x)$ will be reached after a variable delay (Fig.~\ref{fig:f1234}B). 

With some modification, the same transformation applies to enzymatic reactions. The kinetic rate parameters are {\small $Kd ={koff\over{kon}}$} (for $\leftrightarrow$) and $k_{cat}$ (for $\rightarrow$).

\[ [A]+[E_1] \leftrightarrow [AE_1] \rightarrow [A^*] + [E_1]\]

Here it is required that the enzymatic reaction is reversible, i.e. a reaction
\[[A^*] \rightarrow [A] \]
exists. (For instance,
\[ [A^*]+[E_2] \leftrightarrow [A^*E_2] \rightarrow [A] + [E_2]\]
is a reaction that reverses [A*]. ) The differential equations, with $k_{cat2}$ for $[A^*] \rightarrow [A]$, are:

\[ dxdt(A) =  k_{\mbox{\scriptsize\em off}}[AE_1] - k_{on}[A][E_1]+ k_{cat2}[A*]\]
\[ dxdt(E_1) =  (k_{\mbox{\scriptsize\em off}} + k_{cat})[AE_1] - k_{on}[A]*[E_1]\]
\[ dxdt(A*) =  k_{cat}[AE_1]- k_{cat2}[A*]\]
\[ dxdt(AE_1) =  k_{on}[A][E_1] - (k_{\mbox{\scriptsize\em off}} + k_{cat})[AE_1]\]

Given concentrations for $E_1$, A and kinetic rate parameters $Kd$, $k_{cat}$ and 
$k_{cat2}$, we may now derive a function with
$x$ as the source species $[E_1]$ and $y$ as the target species $[A*]$
(Fig.~\ref{fig:f1234}C).

In both cases the resulting curve can be fitted by a saturating hyperbolic function.

\[y=f(x)= y_{max} -({y_{max}-y_{min}\over 1+ ({x\over C})^n}) \] 
Here $y_{min}$, the baseline concentration, is usually set to 0.

If we choose [A] as the target of [E1], we get a negative slope psf. 
\[y=f(x)= y_{min} - ({y_{min}-y_{max}\over 1+ ({x\over C})^n}) \]

We call this function the {\bf elementary protein signaling function} or elementary psf.
This function is somewhat related to a Hill equation \cite{Kraeutler2010,Yu2010}. A Hill equation is a function fitted to an experimental measurement to derive a dose-response relationship, comparable to the psf. The Hill equation allows to calculate a fractional concentration $\theta$ for the target (e.g. a receptor-ligand complex) from the source concentration $[L]$, given $Kd$, and fitting a parameter $n$ for the steepness of the curve.
\[ \theta = {[L]^n \over {Kd + [L]^n}}\]

The concentration of the other compound of the complex is not used (assumed large), and the absolute magnitude of the target is not calculated. An equivalent for enzymatic reactions is 
not defined. The parameter $n$ allows to measure the effect of competing binding reactions (n=1 if none are present),
which in our terminology translates into a systemic psf with multiple binding partners for a single target compound. Systemic psfs are a more general concept than Hill equations, but they relate to the same type of data, namely dose-response functions in steady-state.

We have seen that signal transmission strength is uniformly characterized by saturating hyperbolic functions. 
This means that it is highest for low $x$ and diminishes as $x$ increases (Fig.~\ref{fig:f1234}D). 
For instance, in Fig.~\ref{fig:f1234}D, a 100\% signal increase leads 
to 100\%,  
18\% or only 6\% increase in the target depending on the source concentration.  
For enzymatic reactions, absolute concentration changes have different effects for sources and targets of a signaling interaction. Signal transmission strength depends on the absolute concentration of the source, the target concentration is irrelevant. This is an important observation, since protein signaling systems are subject to long-term regulation of concentrations. 
In the context of disease states or other sources of protein expression up-/downregulation, independence of transmission strength from target concentration may be an important conservative property.

Signal transmission is strongest if a source species is expressed at a low concentration. We need to bear in mind however, that reaction velocity operates inversely to signal transmission strength: a low source concentration means a slow reaction 
(Fig.~\ref{fig:f1234}B).
A functioning signaling system would therefore have to use an intermediate range to
maximize 
signal transmission within time constraints. Our analysis opens a new way of analysis for a signaling system: Optimization techniques could find a best source range for both time and signal transmission constraints.


\subsection*{Systemic PSF Analysis}
\label{systemic1}
A source-target psf can be derived for any pair of species in a complex biochemical reaction system. 
For a complex system, or set of equations, we define a set of input nodes, and compute the output values for each possible input configuration. The analysis gives us the 
output concentration range (notwithstanding transients in a dynamic context, s. below) for 
each species, as well as a (fitted) function, or matrix of input-output correspondences. A biochemical reaction will produce a different psf, when it is elementary or when it is embedded in a context, where the participants of the elementary 
reaction also participate in other reactions. This is true for both protein complex formation and enzymatic reactions.  
We therefore call source-target functions 'systemic psfs', when they 
are derived from the context of a specific signaling system.

We have provided this analysis for the example system. We show the concentration ranges and the signal transmission functions for the whole system \cite{Lindskog2006} as a
weighted dynamic graph for Da as a single input (Fig.~\ref{fig:KIMBL_WGs_VDa_Cahi}). We label each species node with its concentration range, determine source and target species nodes for each reaction node, and provide fits for the systemic psf, the
transfer function that characterizes each reaction.   
We see from (Fig.~\ref{fig:KIMBL_WGs_VDa_Cahi}) that a number of systemic psfs can be fitted well with a linear function ($y=mx+b$), showing that systemic psfs sometimes consist only of a short section from a full mapping of concentration values. Also, many species have only small concentration ranges, which means they don't have much response to Da input.

It is an obvious advantage of the psf analysis that we are able to dissect the complex system and extract local properties, such as concentration ranges of individual species, and transfer functions for individual reactions under input stimulation. This allows to critically analyze a model, compare these properties with biological data, and adjust or improve the model in a detailed manner.

\begin{figure}[!ht] 
\centerline{\includegraphics[width=0.9\textwidth]{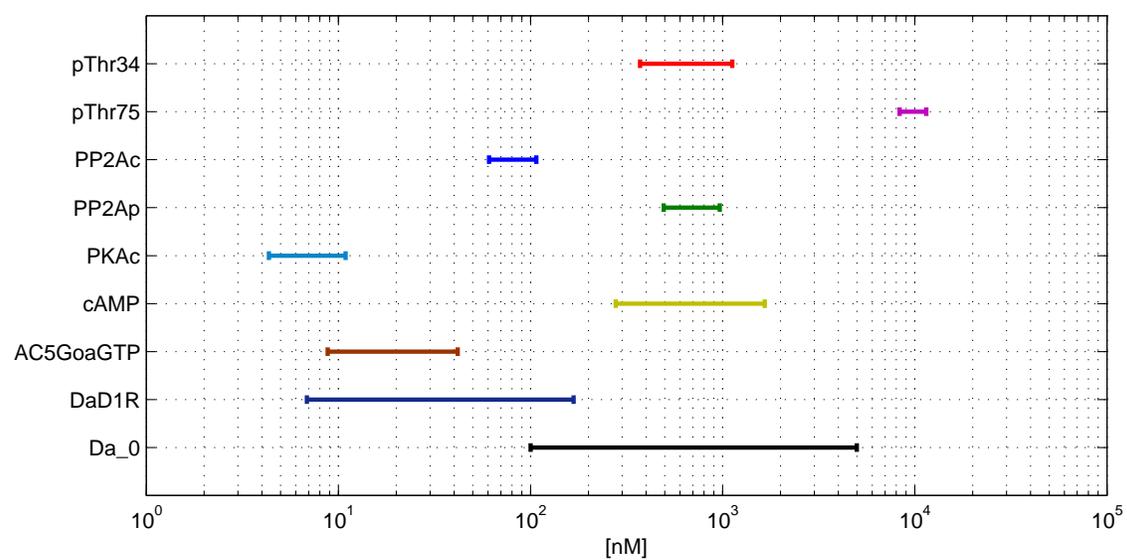}}
  \caption{{\bf Ranges of Concentrations in Response to Input (Da).} For a number of relevant target species from the biological model, the range of concentrations for input (Da) from 100nM to 5$\mu$M is shown (Ca - high, 8$\mu$M).}
  \label{fig:KIMBL_conc_VDa_Cahi}
\end{figure}

In Fig.~\ref{fig:KIMBL_conc_VDa_Cahi}, the concentration ranges for some target species are given. 
We see, for instance, that among DARPP-32 phosphorylation variants, pThr75 is always more abundant that pThr34, by an order of magnitude. This is an example of a high-level property, which could be related to biological data. As another example, we notice that the active receptor conformation (Da-D1R) remains below 160nM even under stimulation with $1 \mu M$ Da and more. With a D1R total concentration of 500nM, we could adjust the ligand binding coefficient to produce more or less active receptors. Finally, the analysis shows a very low maximal PKAc level ($12nM$) 
in spite of a total PKA concentration of $1.2 \mu M$. 
In the original model \cite{Lindskog2006}, blind parameter adjustment has probably generated a very low level of PKAc in order to achieve high signal transmission for phosphorylation of the target species pThr34, which is experimentally required, but which could be achieved in other ways (e.g. PP2B) as well. 

With our analysis, properties of individual species become apparent, and they can be compared to biological data, tested and adjusted on a localized basis. Even more interestingly, we could look for principles of 'rational system design', for instance question the transmission of a seven-fold increase of cAMP in the $\mu M$ range to a maximal three-fold increase in the 10's of nM for PKAc, 
and analyze given biological systems from this perspective. 

\begin{figure}[!ht] 
\centerline{\includegraphics[width=0.8\textwidth]{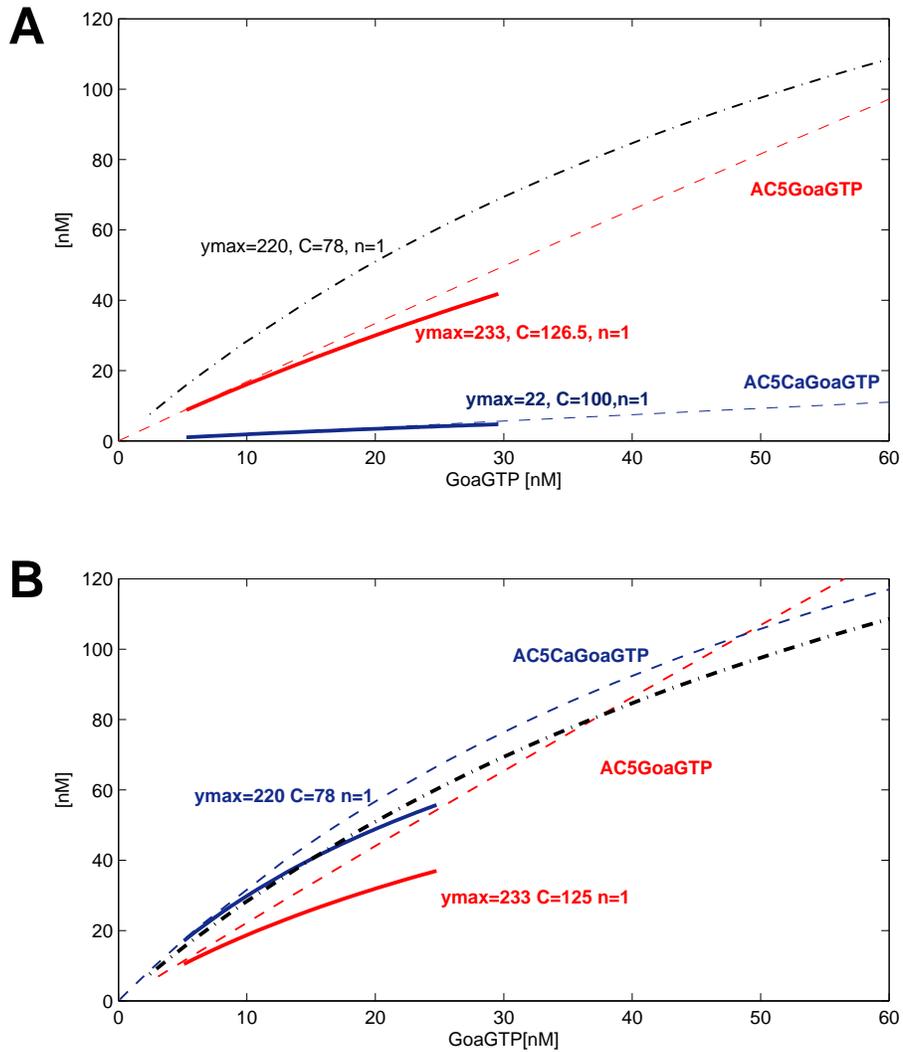}}
  \caption{{\bf Local adjustment of a transfer function.} 
{\bf A.} Elementary (dashed) and systemic (continuous) psfs for two targets of GoaGTP. A new systemic psf for GoaGTP-AC5CaGoaGTP is defined by functional parameter adjustment (black line). 
{\bf B.} Elementary parameters were changed to match the new systemic psf. For AC5CaGoaGTP, $k_{on}$=0.0192, $k_{\mbox{\scriptsize\em off}}$=25 was adjusted
to $k_{on}$=0.022, $k_{\mbox{\scriptsize\em off}}$=1.5, for AC5GoaGTP, $k_{on}$=0.0385, $k_{\mbox{\scriptsize\em off}}$=50 was adjusted to $k_{on}$=0.0495, $k_{\mbox{\scriptsize\em off}}$=48.5.}
  \label{elem-sys}
\end{figure}

In addition to the concentration ranges, we also have access to the functional mapping between species in the model. The systemic psfs, like the elementary psfs, are stored as vectors, which are matched by functional parameters. The advantage of the psf analysis is that we can probe a complex system on a single reaction level because the influence of the cellular context is encoded in the systemic psf. Thus we can compare the elementary psf with its transformation as a systemic psf for individual reactions.
Fig.~\ref{elem-sys}A shows elementary and systemic psfs for G-protein activation of AC5 and the calcium-activated complex AC5Ca. We see that the systemic psfs are somewhat deflected, compared to the elementary psf, which is what we expect from the parallel activation of AC5Ca and AC5 by the same species. We may specify a desired psf using only functional parameters, and adjust elementary parameters to match the psf (Fig.~\ref{elem-sys}B).  A local change to the $Kd$ binding coefficient between AC5Ca and GoaGTP allows a change in the systemic function. Since other systemic psfs may be affected by such a change - this can be detected by re-computing the weighted dynamic graph - more adjustments of elementary rate parameters may be indicated, possibly by an iterative process 
(cf.~\cite{Maslov2007b}). 

\begin{figure}[!ht] 
\centerline{\includegraphics[width=0.95\textwidth]{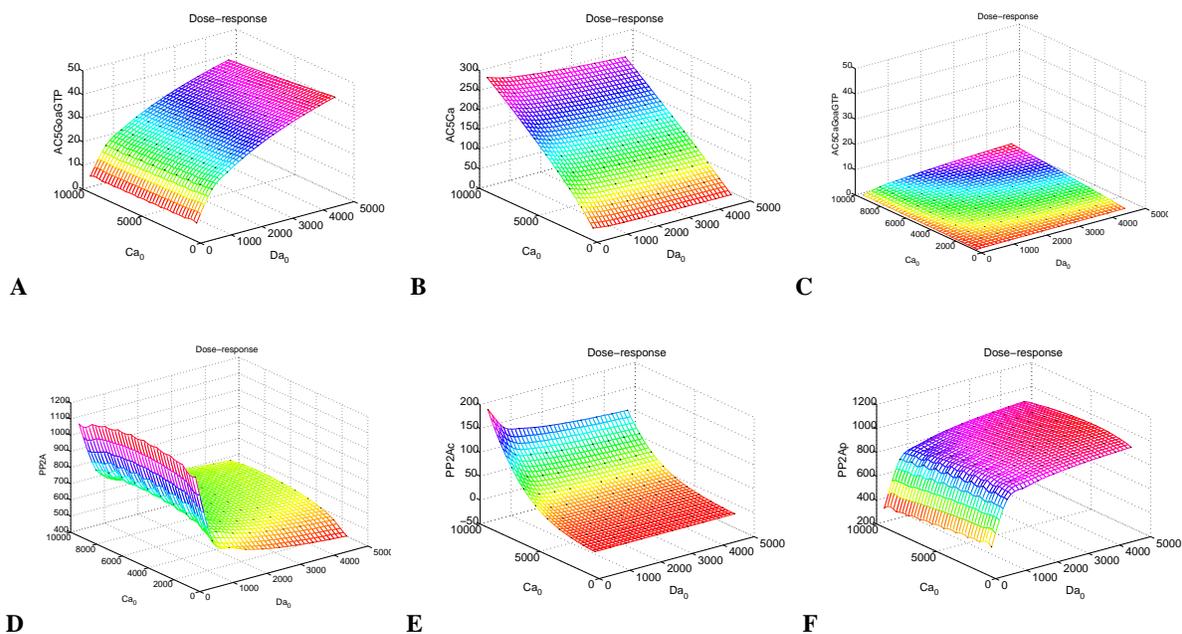}}
  \caption{{\bf Dependence of target species on Ca and Da inputs.} The 2D psf shows species which segregate to only one of the input pathways ({\bf A}, {\bf B}, {\bf F}), a species which remains unresponsive ({\bf C}), and species which show some signal integration in their input constellation ({\bf D}, {\bf E}).}
  \label{fig:KIMBL_2D_AC5GoaGTP}
\end{figure}

Sampling for multiple 
inputs yields a transfer function matrix, which can be analyzed for dependence of the target concentration on each input separately. This can be done by standard matrix analysis such as principal component analysis (PCA).
For our example, we show how species which are poised to integrate signals from two different sources do this under the numeric conditions (Fig.~\ref{fig:KIMBL_2D_AC5GoaGTP}). 
For cAMP production (AC5), we find that AC5GoaGTP is dependent only on Da
(Fig.~\ref{fig:KIMBL_2D_AC5GoaGTP}A),  
AC5Ca is only dependent on Ca (Fig.~\ref{fig:KIMBL_2D_AC5GoaGTP}B), while AC5CaGoaGTP is almost not activated at all (Fig.~\ref{fig:KIMBL_2D_AC5GoaGTP}C).
Even though a link of reactions (Ca-AC5Ca-AC5CaGoaGTP-cAMP) exists, signal integration of Ca and Da on AC5 fails because of the weak transmission from GoaGTP to AC5Ca. 
Signal integration between Ca and Da occurs for cAMP degradation by calcium-dependent calmodulin regulation of PDE1. 
PP2A with the two variants (calcium-activated) PP2Ac and (PKAc-activated) PP2Ap is another potential source of signal integration 
(Fig.~\ref{fig:KIMBL_2D_AC5GoaGTP}D,E,F). 
The psf analysis shows when signal integration occurs (here: Da having influence on PP2Ac),
and when this effect is negligible (here: Ca not having influence on PP2Ap). This may now be studied for correspondence with the biological situation.  These results emphasize the necessity for numeric analysis of input-dependence, beyond the mere existence of links.


\subsection*{Systemic Delay and State-change Dynamics}
\label{delay}
We would like to be able to use systemic psfs with their simple and transparent mathematical structure for dynamical simulations. This allows direct experimental testing and fitting by time series measurements  beyond dose-response relationships. In order to do this, we need to compute the systemic delays, i.e. the reaction time until a steady state is formed. Then we can build a state-change dynamical model from systemic psfs alone, using the appropriate delays for the input and the update of a system state. 

Systemic delays depend on the absolute size of the signal and also the direction (increase/decrease) of signaling. Delays for species in the example system in response to input are shown in Table~\ref{delay-table}. For the computation of target concentrations, we only need a ratio such as ${kon \over \mbox{\scriptsize\em koff}}=Kd$ (binding coefficient) or $kcat \over kcat2$ for forward and backward enzymatic reactions. For the delays, the difference between $k_{on}$ and $k_{\mbox{\scriptsize\em off}}$ or $k_{cat}$ and $k_{cat2}$ defines reaction times for synthesis and degradation. Therefore, delay computations are fairly complex, but the results are often
within a fairly narrow range for each reaction (Table~\ref{delay-table}). 
For discrete state-change simulations we may use maximal delays for each species.

\begin{table}
\begin{tabular}{l|l|l|l|l|}
Species & {\small $0.06 \rightarrow 0.5\mu M$ }& {\small $0.06 \rightarrow 4.5 \mu M$} &{\small $0.5 \rightarrow 0.06 \mu M$}& {\small $4.5 \rightarrow 0.06 \mu M$}\\
\hline

DaD1R      &  {\bf 3.6}       & {\bf 0.8 }&10.1&10\\
AC5GoaGTP  & {\bf 7.8}  	& {\bf 2.4}&19&20\\
AC5CaGoaGTP& {\bf 8.1}         &	{\bf 2.4}    &19.5&26\\
cAMP       & {\bf 7.8}   &	{\bf 3.1}    &18.1&18\\
PKAc       & {\bf 251}   	 	&{\bf 164}  &347& 300\\
pThr34     & {\bf 288}     &{\bf 200}  &369&307\\
PP2Ap      & {\bf 511}  	&{\bf 387} &706&683\\
PP2Ac      & {\bf 600}     	&{\bf 483} &756&717\\
pThr75     & {\bf 493}     &{\bf 359}&756& 755
\end{tabular}
\caption{Delays:
For near instantaneous input signal Da at shown concentrations, the table shows delays (in s) to reach EC5/EC95 for target species. Decrease is usually slower than increase, due to the asymmetry of $k_{on}$/$k_{\mbox{\scriptsize\em off}}$ binding parameters. Delay times are sensitive to the absolute size of the signal, with delays being faster for larger signals.
}
\label{delay-table}
\end{table}

\begin{figure}[!ht] 
\centerline{\includegraphics[width=0.95\textwidth]{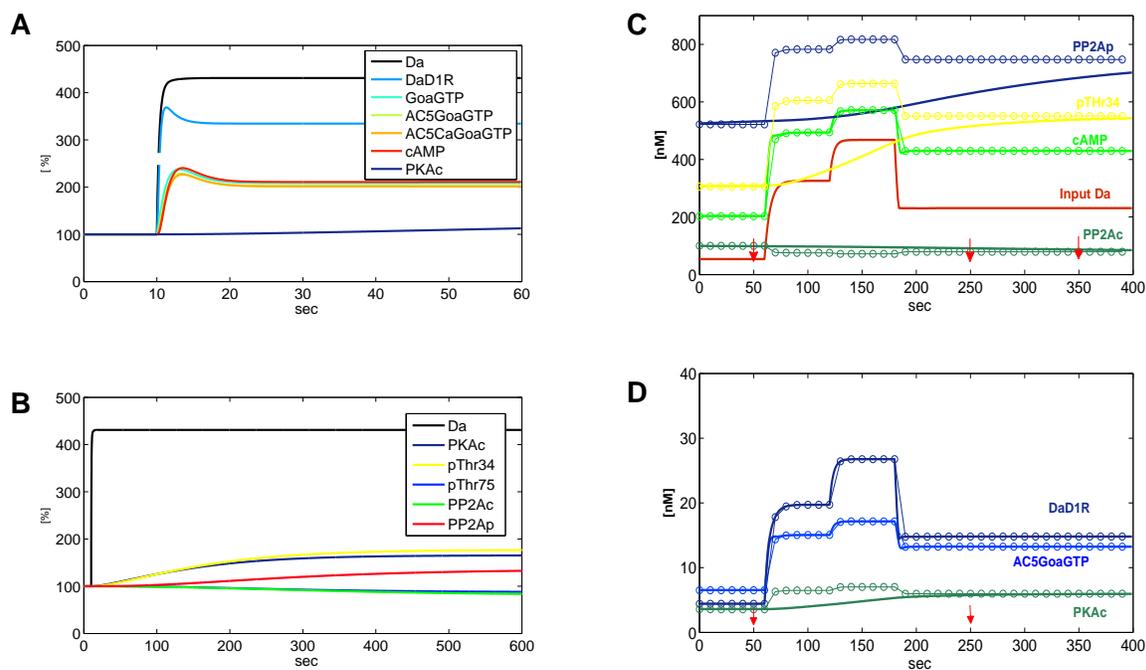}}
\caption{
{\bf Dynamics of target species to Da input.} {\bf A.} Fast species ($<$10s) have transients. The psf approximation only calculates the steady-state value. {\bf B.} For slow species no transients are apparent. It may take several minutes to reach steady state.
{\bf C, D.} Dots mark systemic psf values, thick lines are continuous dynamical simulations by differential equations, thin lines are interpolations between 10s psf values. Red arrows mark time points for interpolations for slow species like PP2Ac, PP2Ap, pThr34.}
  \label{fig:f910}
\end{figure}

From a biological perspective, this table provides an important test on the validity of the model. In many cases, systemic delays can be measured. For instance, the delay for PKAc at 150-250s rather than 30-60s, as measured in \cite{Woo2003} (cf.~\cite{Kim2010}), seems large and may be an indication for a revision of the underlying parameters.  From the theoretical perspective, this system seems to operate on separate time scales:  1-10s, 150-300s and 450-600s. Such a separation of reactions by their characteristic delay times is interesting, since it could lead to simulation models with different discrete time scales. Here we may calculate psf values for fast species with 10s time resolution, for intermediate species with 300s time resolution, and for slow species with 600s time resolution, i.e. system update time for state changes (Fig.~\ref{fig:f910}A,B). It is an empirical question, whether separate time scales rather than a continuum of delay values will prove to be an organizing principle in protein signaling 
systems \cite{Purvis2009}. A general study, for instance, using models from the BioModels Database \cite{Li2010}, might give answers to this question. 
Time scale separation may provide a conservative property of a signaling system against fluctuations of concentrations. 
If total concentrations in the system change, e.g. by protein expression up- or down regulation, miRNA interaction, or diffusional processes across compartments, the relevant interactions will continue within each time slice. Concerted regulation of protein expression levels may set a clock for the rapidity of signal transduction.

Systemic dynamics, in contrast to elementary reaction dynamics, need not follow a hyperbolic curve. If there are feedbacks in the system, the dynamics may contain transients, i.e. the concentration may be higher or lower before it settles into its steady-state value \cite{Theis2011}. The dynamic response of target species to input are shown in Fig.~\ref{fig:f910}A,B. For a species without a transient response, the actual value of a species at a shorter delay is always bounded by the steady-state value, and all possible concentrations in a 
continuous-time dynamical system are bounded by the psf concentration range 
\cite{Purvis2009}. However, if there are transients, a psf-based dynamical simulation will miss these transients and plot a simplified trajectory.
This means that results from a psf analysis
with slow inputs cannot be extrapolated to much faster input dynamics - in contrast to continuous-timed dynamical systems where arbitrary time units can be chosen. This restriction may capture a biological reality: steady-state behavior provides the framework and may operate according to rules and principles which are separate from the effects of short term fluctuations.

The psf system allows to generate a dynamical system as a sequence 
of states defined by fluctuations of input. Fig.~\ref{fig:f910}C,D shows an overlay of a differential equation simulation and psf state change simulation for a sequence of inputs with 10s duration. Accordingly, the psf approximation is excellent for all species with a delay time of $<10s$. If we plot psf values for slow species at intervals corresponding to their maximal delays, we may linearly interpolate between 
points, and in this case achieve a good approximation of the continuous-time model. 


A psf-based dynamical system is an important tool in order
to generate a time-series simulation from a calculated model system for comparison 
with experimental data. The psf system utilizes parameters which are uniform and have linear error ranges (cf.~Fig.~\ref{fig:f1234}A,C), and therefore should improve interaction of the model with the experimental reality. The psf model will also allow to predict the optimal stimulation times for different inputs such that responses can be measured in steady-state.


\subsection*{Computing Input-dependent Modularity}
\label{modularity}
Since we are able to define signal transmission capacity, we have a tool to investigate modularity of a signaling system. As species saturate or return to basal levels, 
they act as inactive links, i.e. they are stuck at the same concentration value, and cannot transmit further increases or decreases of inputs. We hypothesize that this effect is actually important in many protein signaling systems. We may define an inactive connection as a species node which has only limited (e.g. $<10\%$) signal transmission capacity. The interconnectedness of a system is then proportional to the number of inactive connections, and a module is a part of the system with few or no active connections to the rest of the system. 

\begin{figure}[!ht] 
\centerline{\includegraphics[width=0.8\textwidth]{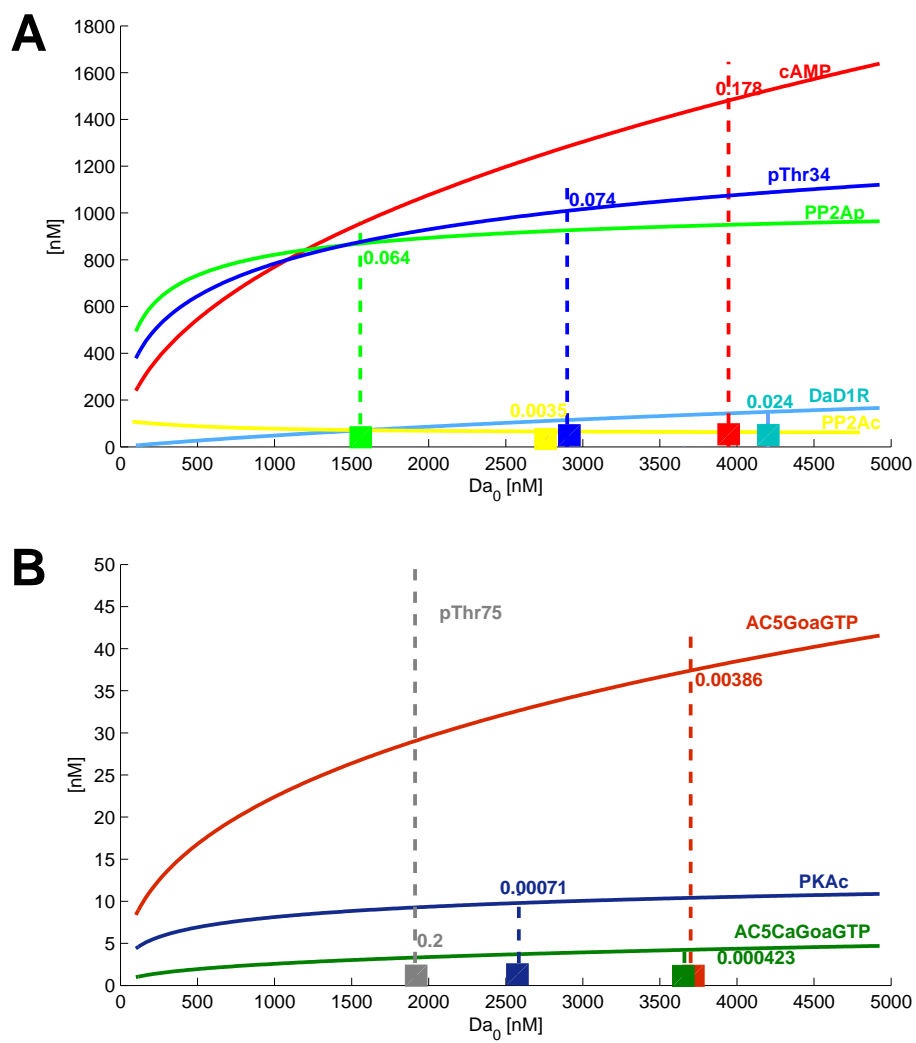}}
  \caption{{\bf Systemic psfs for input to target.} Shown are EC90 and EC10 concentrations as cut-off thresholds for effective signaling by input and the slope values at threshold.}
  \label{fig:KIMBL_psf_EC90_1}
\end{figure}

In the following we discuss the activation/inactivation of links with respect to input 
increases. I.e. given a certain level of extracellular signaling, what happens if this level is raised and then kept at the higher level for some time, sufficient for the system to settle into a new steady state? Which species nodes will respond to the increase and transmit it to downstream targets, and which species will become saturated and only 
respond with their saturated value? It is also clear that species which have become saturated (inactive) will not respond to fast extracellular signals anymore. This analysis is therefore useful both for the steady-state context and for 
understanding fast input fluctuations.
There is an input level for each target species, where the species ceases to be responsive to further input increases. If that input level has been reached, the species can be considered to have become an inactive connection, i.e. a node which does not transmit signals. We may define systemic psfs for 
'input-target psfs' from the input to any target species (Fig.~\ref{fig:KIMBL_psf_EC90_1}).
We notice how number of steps in the computation of a species concentration relates to a lower cut-off value for signal transmission (e.g. DaD1R, AC5GoaGTP, cAMP vs. PKAc, pThr34, pThr75). In other words, earlier steps in the sequence saturate at a higher input level than later steps. Two parallel targets (pThr34, PP2Ap) of an intermediate step (PKAc) may saturate at very different input levels ($\sim 3 \mu M$ for pThr34 vs. $1.5 \mu M$ for PP2Ap). This mechanism demonstrates the effect of a sequence of saturating functions, and constitutes a general principle in the construction of a signal transduction system.

\begin{figure}[!ht] 
\centerline{\includegraphics[width=0.99\textwidth]{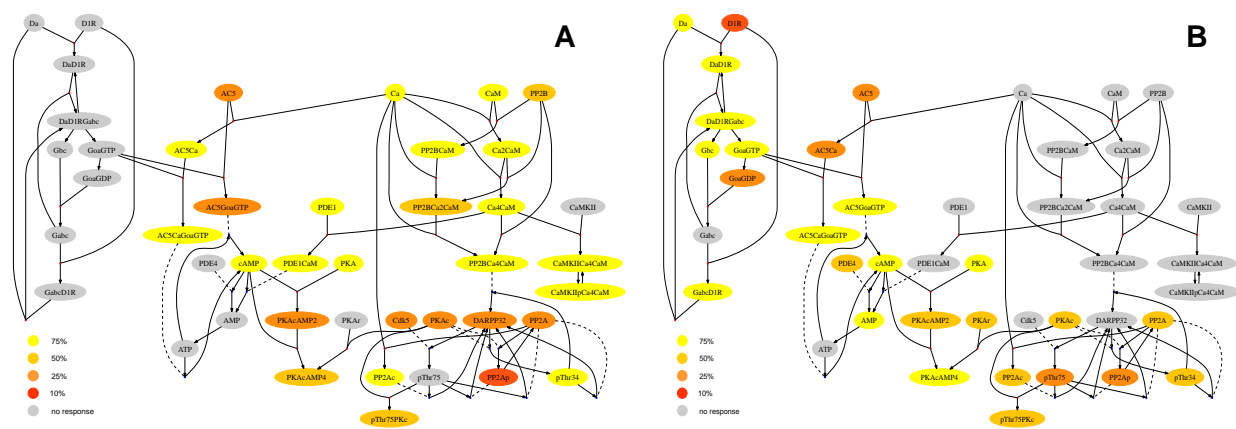}}
\caption{{\bf Modularity in signal transmission.} Species nodes are colored according to their saturation/depletion status (EC90/EC10) in response to percentage of input. 
{\bf A.} Input is Ca 
(100nM...10$\mu M$), Da is set at 5$\mu M$. The figure shows a large number of species which have no response (less than 10\%) to Ca input, two separate pathways are apparent. Species inactivate at low input levels in the 'integration zone'.
{\bf B.} Input is Da (20nM...5$\mu M$), Ca is set at 8$\mu M$. The graph shows input-dependent inactivation of links. Individual effects can be studied. For instance, PDE4 (in contrast to PDE1, PDE1CaM) shows responsivity to Da input because of a larger complex formation with cAMP, which subtracts from the enzyme concentration.}  
  \label{fig:f1213}
\end{figure}

The model allows to study whether a node responds to a specific input with any change of steady-state. 
In Fig.~\ref{fig:f1213}
and supplemental figures~\ref{suppl:f1}, \ref{suppl:f2}, we show modularization of the system under various Da and Ca input conditions. With Da input
and high Ca (Figure~\ref{fig:f1213}B), species
which are proximal to Da input, the receptor-ligand complex DaD1R, the signaling complex through G proteins and adenylyl cyclase AC5, as well as cAMP, are most responsive to Da over a large range of input. Species in the 'integration zone' between Da and Ca, such as DARPP-32, PP2A cease responding to Da increases at lower levels and become inactive links at higher levels. Species in the system with no significant change in concentration at any input level are for the most part proximal to Ca instead of Da, and thus highlight modularity among pathways. In this case, we see that Da inputs are transmitted to distant targets only up to an intermediate range and that there are a significant number of species which do not react to Da at all.  Above that range, even though closely coupled targets still respond to the input, the signal increase is not registered beyond cAMP production and synthesis. 
With Ca low (supplemental Figure~\ref{suppl:f1}) there is widespread responsivity to Da up to 1-2$\mu$M and only a few of the calcium-related species do not respond at all.  The Da signal is therefore able to influence the calcium-related pathway, provided calcium is low.


When Ca is the input (Fig.~\ref{fig:f1213}A), the calcium-responsive proteins like calmodulin, CaMKII, calcium-activated PP2B (calcineurin) transmit signals, while the GPCR pathway remains almost completely unresponsive. There is some signal integration with calmodulin-activated PDE1 for cAMP. Other than that, we see that PP2Ac and the pThr34 variant of DARPP-32 respond strongest to calcium while PP2Ap and the pThr75 variant of DARPP-32 have less or no responsiveness to calcium. With Da low, as in supplemental Figure~\ref{suppl:f2}, again, most of the GPCR-related species do not respond to Ca at all, or cease responding at 10-25\% of maximal Ca (1-4$\mu$M). However Ca-related species like calmodulin, CamkII, PP2B remain responsive.  The difference between the high and low Da condition is small. Here the few existing links from Ca to GPCR (such as Ca regulation of AC5) are not strong enough to influence the GPCR pathway significantly in any condition. In this case, it is not just the saturation that matters, the input signal has limited reach in influencing distant species in general.

We are analyzing a biological system with two 'pathways', the cAMP and the calcium pathway, which are cross-linked in a convergence zone of species which are influenced by both pathways. It is remarkable how clearly three different modules 
appear: the GPCR/cAMP pathway, the Ca pathway, and the signal integration zone.

There is also a general observation to be made about signal transmission in a protein signaling system: Signal integration is strongest when inputs are low. This is a direct consequence of the effect of coupling saturating nodes. It means that there are few saturating species in the system which impose modularity, and signals spread further.
Widespread interconnectedness is only possible at low input levels. Because activation curves for biochemical reactions are mostly uniform hyperbolic (saturating), a stronger modularization with many inactive links results from higher input levels. This may also have implications on transient responses. For instance, phosphatase and kinase response often differs with kinases being saturated only at higher input levels. If and where that is the case, we may observe transient responses that only reflect kinase activity, since phosphatases are saturated and do not participate in signaling, they do not add or subtract and in this way obscure the kinase signal. This shows that the
steady-state input-response system may work well as the framework wherein fluctuating signaling operates.


\section*{Discussion}
Continuous dynamical systems - systems that use change over time as a system primitive - are notoriously difficult to analyze and may not be the best choice of a tool for signal transduction systems of moderate or large size. Steady-state matrix computations are simple and fast, scale well to very large sizes, and offer multiple opportunities for analysis. By calculating transfer functions from a systemic dynamical model, we also gain the opportunity to extract and analyze parts of the system. This may help in creating re-usable system parts.
This paper demonstrated a transformation of a mass-action kinetic biochemical reaction model implemented by a set of differential equations into an input-response transfer function model. 
The transformation is done by calculating steady-state concentrations for each species in response to a range of input values, and then analyzing the resulting vectors (matrices) as the basis of a transfer function ('psf'). For small, toy-like networks of few components, such dose-response relationships have been investigated by \cite{Tyson2003}.
They analyse kinetic models in the same way, however, they don't make a distinction between parameters in elementary interactions, and the actual parameters in a systemic context. They also do not address the question of temporal embedding of the dose-response relations. In our approach, we use dose-response functions, similar to Hill functions, but we are extending the concept. By using rectangular signals, i.e. constant signaling levels, we calculate the response as the 
steady-state value. In addition, we calculate the time to steady-state from the underlying dynamical model. 
Only because of this additional computation can one attempt to create a 
discrete dynamical model -- something that is not within the purview of a Hill equation model. Hill equations attempt to fit or create systemic parameters, which are different from elementary parameters, i.e. they do recognize the dependence of the transfer function on the systemic context. The psf model is an approach of making Hill equations (dose-response relations) work in large-scale modeling.

There are numerous attempts to simplify dynamical models in the temporal domain by creating hybrid models (e.g. \cite{Eriksson2009}). Sometimes slow interactions are regarded as constant, and only fast interactions are dynamically modeled. Sometimes, fast interactions are replaced by time-independent values and only slow interactions are dynamically modeled (many examples, for instance \cite{Saucerman2003}). There are also 
attempts to replace an ODE model by a delay-differential equation (DDE) model, i.e. 
to compute and use explicit time delays and eliminate many intermediate species, simplifying the size of the model \cite{Srividhya2007}.

In our approach, time and concentration are regarded as separate, which makes it different from any hybrid approach.
The main restriction is the assumption of a constant signaling level over periods of time sufficient to induce steady-state. The approach is therefore best suited to check the limiting conditions of a dynamical model, e.g. in drug development applications, where multiple dose-response relations derived from the model can be cross-checked and used to locally improve the model. 
However, the simple, atemporal transfer functions can also be applied directly: (a) in experimental settings where signal duration can be controlled, and (b) in physiological settings, when fluctuations occur around different mean values, and we model the step changes for the mean extracellular signaling level, but not the short-term fluctuations. 
It could be shown that psfs are sufficient to create discrete time dynamical models, with certain restrictions on fast dynamical inputs below the time resolution of the system. The primary focus of the analysis was on single-input systems, where signaling functions can be matched by parameterized hyperbolic functions. We showed how the analysis can be extended to multiple-input systems by computing and analyzing psf transfer matrices. 
The psf functions allow to cut through the complexity of the model and examine interactions in a localized way. If continuous dynamical models are being used with the goal of adequately simulating cellular processes, this kind of analysis is an indispensable tool to check for the consequences of the modeling assumptions in terms of dose-response relationships. The analytical tools of the psf approach may also be used to systematically investigate the effect of localized changes to the system 
\cite{Maslov2007a,Maslov2007b},
and to offer local corrections to the complex system in a transparent way.

The extraction of input-target psfs with characteristic hyperbolic saturating properties allows to determine input level dependent inactivation of a species node. Accordingly, we can define the limits of signal transmission by the distribution of inactive nodes. In the example system, a strong distinction of calcium- and cAMP-dependent pathways and a signal integration zone were revealed. 
We could see that at low levels of input, widespread interactions are possible, while at higher levels of input, many species enter into a state of a constant function value and become inactive links. 
This corresponds to biological results and expectations, and provides a  
foundation for the concept of pathways in signal transduction.  
For transient responses, the
steady-state input-response system provides the boundary values to which the system eventually settles. This is especially interesting if we have saturating response levels, which are unresponsive to further signaling (e.g. phosphatases), and allow transient signals (e.g. kinases) to emerge. 

Signal transduction may also be analyzed from the perspective of rational system design. Such work is still in its infancy. We may for instance investigate the effect of negative feedback links on concentration ranges and times to steady state.
Another question would be the optimization of a signal transduction system for the trade-off between speed and efficacy of signal transmission. With this choice of model, many new questions can be raised, and old problems like parameter dependency, modularity or signal integration can be addressed in a novel way.


\clearpage

\centerline{\includegraphics[width=0.99\textwidth]{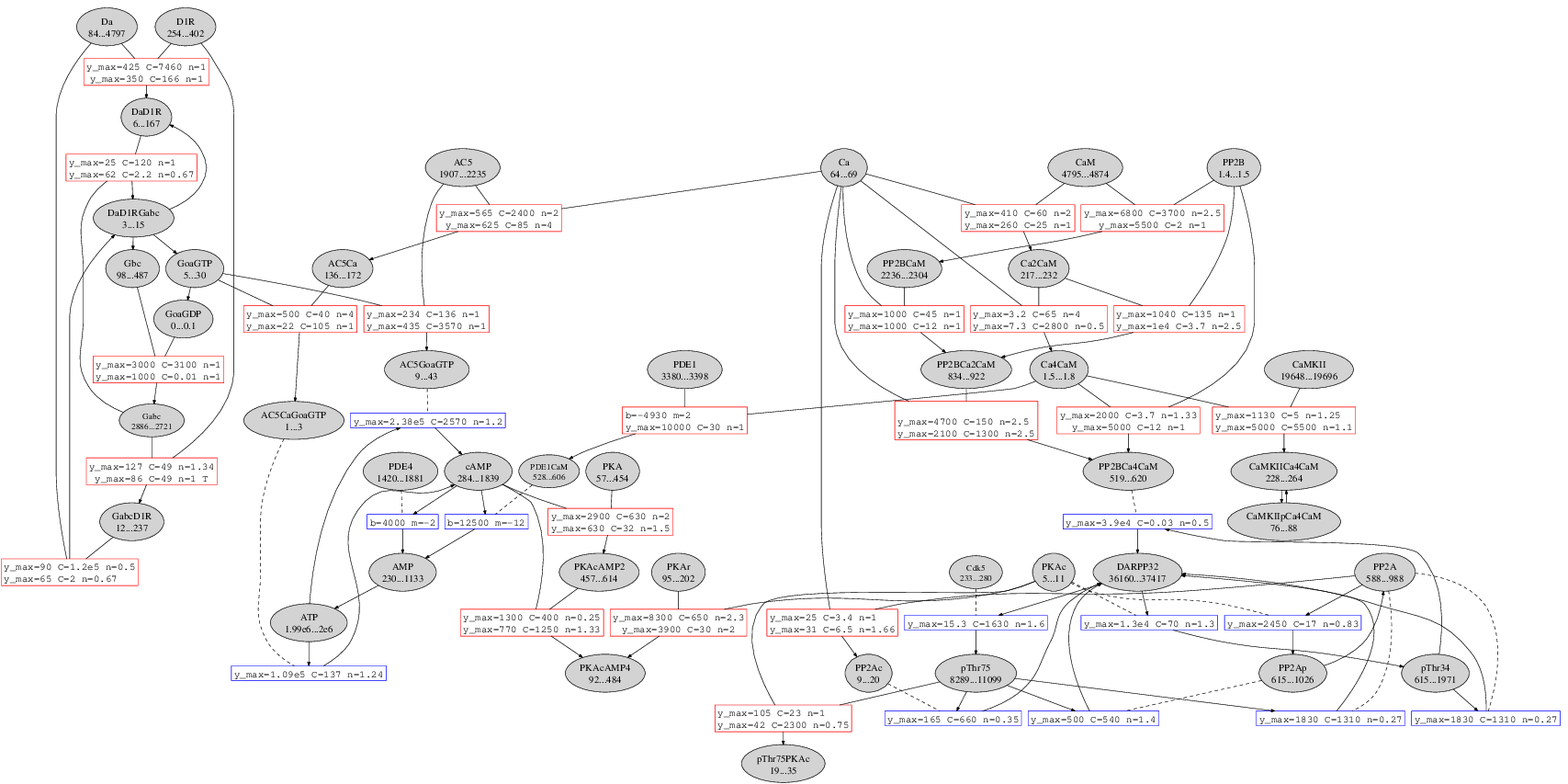}}
\begin{figure}[!ht] 
  \caption{{\bf Weighted Dynamic Network View of the Model System.} Input conditions are Ca=$8 \mu M$ and Da=$60nM-5 \mu M$ (total available concentration). Each species node is labeled with its steady-state concentration range. Reaction nodes are labeled by parameters for a hyperbolic fit, or a linear fit if that was sufficient. Complex formation reactions (red) are labeled for both $[A]\rightarrow[AB]$ and $[B]\rightarrow[AB]$ (left to right), enzymatic reactions (blue) are only labeled for $[E]\rightarrow[A*]$. The low free concentration for Ca (64-69nM) is a feature of this model, where Kds are set in such a way that most Ca is bound to calmodulin, and then to other proteins (cAMKII Kinase, PP2B phosphatase and PDE1), such that all ions are indeed accounted for (some species have 2 or 4 Ca ions bound). The present analysis allows to critically evaluate the effect of elementray kinetic parameters on steady-state properties. }
  \label{fig:KIMBL_WGs_VDa_Cahi}
\end{figure}

\clearpage
\setcounter{figure}{0}
\centerline{\includegraphics[width=0.9\textwidth]{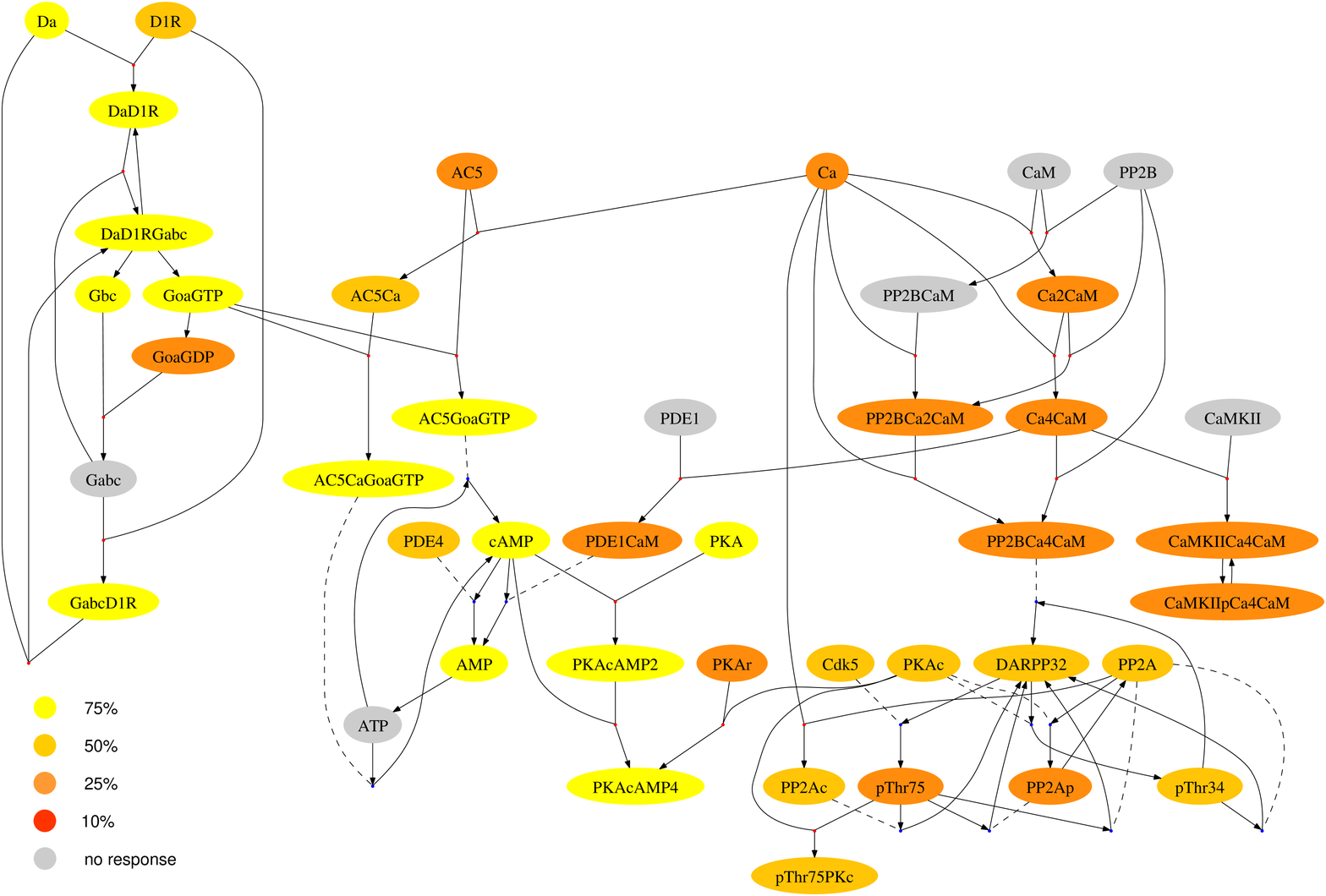}}
\captionsetup[figure]{name={Supplemental Figure}}
\begin{figure}[htb] 
  \caption{{\bf Graph representation of signal transmission.}  Input concentration for Da is $20nM$ \ldots $5\mu M$, for Ca is $100nM$. There is widespread interconnectedness for low inputs.}
  \label{suppl:f1}
\end{figure}

\begin{figure}[htb] 
\centerline{\includegraphics[width=0.9\textwidth]{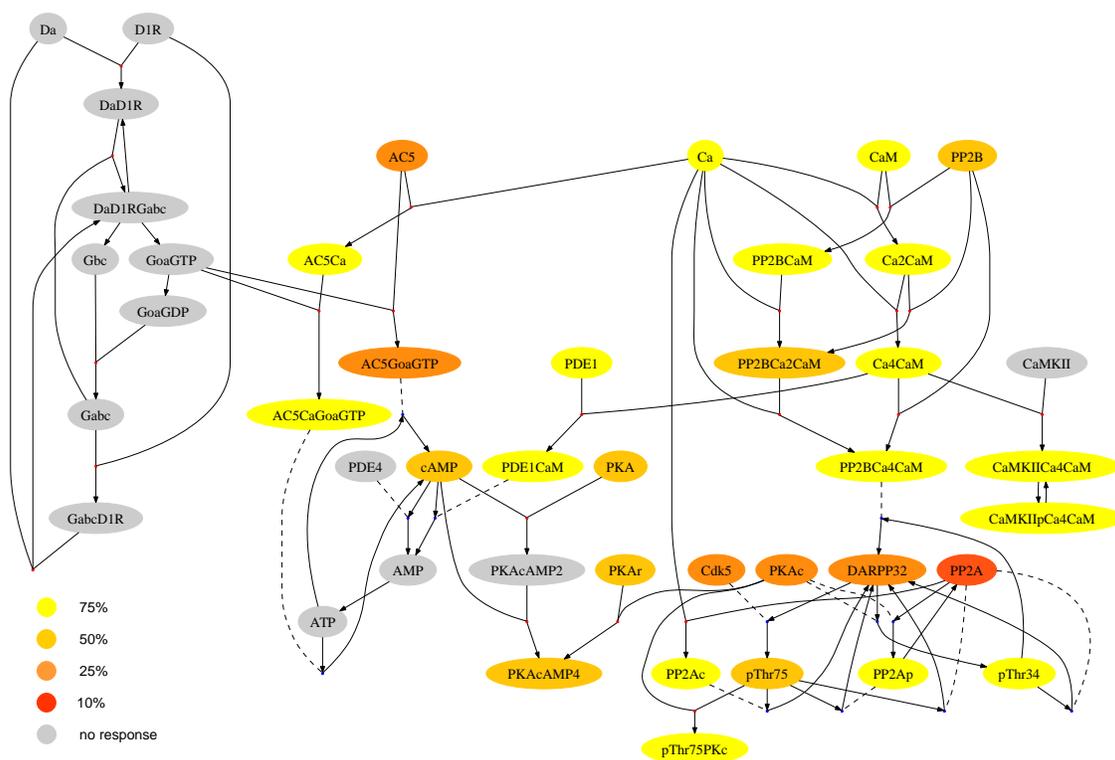}}
  \caption{{\bf Graph representation of signal transmission.} Input concentration for Da is $100 nM$, for Ca is $100 nM$ \ldots $10\mu M$. The Da/GPCR pathway is clearly separated from Ca inputs.}
  \label{suppl:f2}
\end{figure}

\end{document}